\documentclass[a4paper,fleqn]{cas-sc}

\usepackage[numbers,sort]{natbib}
\usepackage{siunitx}
\usepackage{soul}
\usepackage{cancel}
\usepackage{lineno}

\def\tsc#1{\csdef{#1}{\textsc{\lowercase{#1}}\xspace}}
\tsc{WGM}
\tsc{QE}

\begin{document}
\let\WriteBookmarks\relax
\def\floatpagepagefraction{1}
\def\textpagefraction{.001}

\shorttitle{SCALE-TRACK}    

\shortauthors{Schmalfuß et al.}  

\title [mode = title]{SCALE-TRACK: Asynchronous Euler-Lagrange particle tracking on heterogeneous computing architecture}

\affiliation[1]{organization={Leibniz-Institute for Tropospheric Research},
            addressline={Permoserstr. 15}, 
            city={Leipzig},
            citysep={}, 
            postcode={04318}, 
            state={Saxony},
            country={Germany}}

\affiliation[2]{organization={Wikki GmbH},
            addressline={Ziegelbergsweg 68}, 
            city={Wernigerode},
            citysep={}, 
            postcode={38855}, 
            state={Saxony-Anhalt},
            country={Germany}}

\author[1]{Silvio Schmalfuß}[type=editor,
                            orcid=0000-0003-2347-0956,
                            linkedin=silvio-schmalfuss/]
\cormark[1]
\ead{silvio.schmalfuss@tropos.de}
\ead[url]{https://www.tropos.de/en/institute/about-us/employees/silvio-schmalfuss}
\credit{conceptualization (supporting), data curation (equal), formal analysis (equal), funding acquisition (equal), investigation (equal), methodology (equal), project administration (lead), resources (equal), software (equal), validation (equal), visualization (lead), writing – original draft (lead), writing – review \& editing (lead)}

\author[2]{Sergey Lesnik}[orcid=0009-0007-5703-4438, linkedin=sergey-lesnik-67694aa2]
\credit{conceptualization (supporting), data curation (equal), formal analysis (equal), funding acquisition (supporting), investigation (equal), methodology (equal), resources (equal), software (lead), validation (equal), visualization (equal), writing – review \& editing (equal)}

\author[2]{Henrik Rusche}[orcid=0009-0003-3207-8937, linkedin=henrik-rusche-b755b533]
\credit{conceptualization (lead), data curation (equal), formal analysis (equal), funding acquisition (lead), investigation (equal), methodology (lead), project administration (supporting), resources (equal), software (equal), supervision (equal), validation (equal), visualization (supporting), writing – review \& editing (supporting)}

\author[1]{Dennis Niedermeier}[orcid=0000-0002-8265-6235]
\credit{methodology (supporting), supervision (supporting), validation (supporting), writing – review \& editing (supporting)}


\begin{abstract}
Euler-Lagrange (EL) simulations provide a direct and robust framework for modeling disperse multiphase flows. However, they are computationally expensive. While various approaches have attempted to leverage heterogeneous computing architectures, they have encountered scalability limitations. We present SCALE-TRACK, a scalable two-way coupled EL particle tracking algorithm, designed to exploit heterogeneous exascale computing environments. With asynchronous coupling, cache-friendly data structures, and chunk-based partitioning, we address key limitations of existing EL implementations. Validations against an analytical solution and a conventional EL implementation demonstrate the accuracy of the proposed algorithms. On a local workstation, we simulated 1.4 billion particles in a test case featuring a single graphics processing unit (GPU). Scaling runs on an HPC (high-performance computing) cluster show excellent strong and weak scaling, with up to 256 billion particles being tracked on up to 256 GPUs. This represents a significant advancement for EL simulations, enabling high-fidelity simulations on local workstations and pushing the limits on HPC systems. The software is released as open source and is publicly available.
\end{abstract}

\begin{graphicalabstract}
\includegraphics[width=13cm]{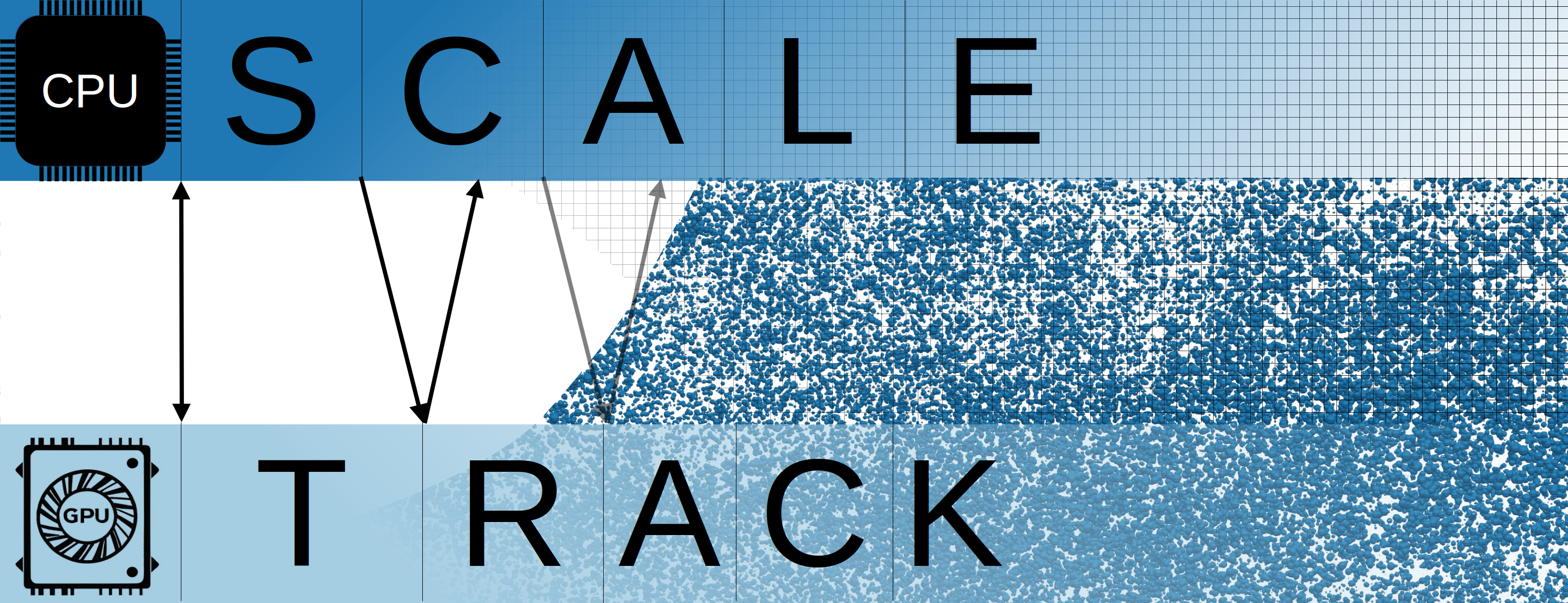}
\end{graphicalabstract}

\begin{highlights}
\item Asynchronous algorithm for two-way coupled Euler-Lagrange simulations
\item Exascale readiness demonstrated for up to 256 billion particles on 256 GPUs
\item Open source code that couples to any suitable CFD solver 
\end{highlights}

\begin{keywords}
 \sep Euler-Lagrange \sep Multiphase flow \sep CFD \sep HPC \sep GPU \sep Exascale 
\end{keywords}

\maketitle

\section{\label{sec:intro}Introduction}

Disperse multiphase flows are ubiquitous in nature (e.g.~clouds~\cite{bodenschatz2010}, sediment transport~\cite{jaiswal2025}, or aerosol propagation~\cite{pohlker2023}), and technology (e.g.~inhalers~\cite{sommerfeld2016,schmalfuss2017}, mixing~\cite{sommerfeld2020,schmalfuss2023}, or internal combustion engines~\cite{blume2019}). Euler-Lagrange (EL) simulations~\cite{crowe1977,niedermeier2020,niedermeier2025} are a common modeling approach, in which the continuous phase is calculated in an Eulerian frame, e.g.~by using the finite volume method, and the disperse phase is handled in a Lagrangian frame by tracking the paths of a multitude of dispersed elements, later called particles. Depending on the volume and mass fraction of these particles, the simulations may be one-way, two-way, or four-way coupled~\cite{michaelides2017}. If the influence of particles on the continuous phase is negligible, it is sufficient to use one-way coupling, i.e.~to only consider the influence of the continuous phase on the disperse phase. As soon as there is a significant feedback from the disperse to the continuous phase, two-way coupling becomes necessary. When interactions between particles, i.e.~inter-particle collisions, become relevant, the simulation needs to be four-way coupled. The focus of this work is on two-way coupled simulations, as for example used in cloud simulations~\cite{chen2025}.

One advantage of the EL approach is the comparatively simple description of the disperse phase – much simpler than in mesh-based methods that treat the disperse phase like a continuum, as for example the Euler-Euler method~\cite{sokolichin1997} or the several variants of the method of moments~\cite{marchisio2005}. Another advantage is that this method can serve as the basis for deriving empirical correlations for reduced models, as it provides direct access to the basic properties of the particles. The simplicity and usefulness of this method comes at the cost of a high need for computational resources, often needing significantly more computational time than reduced models.

Nonetheless, owing to the steady increase in computational power, the EL method has been applied increasingly over the last few decades~\cite{michaelides2017}. In addition to increased computational power, there is a trend towards heterogeneous computing architecture in HPC (high performance computing) clusters~\cite{kemmler2025,muralikrishnan2024}, i.e.~central processing units (CPUs) and graphics processing units (GPUs) being available at the same compute nodes. Because each Lagrangian particle is tracked independently but follows the same computational procedure, the method exhibits massive parallelism. This SIMD-style (Single Instruction, Multiple Data) computation is where GPUs deliver their greatest performance gains. It therefore makes sense to compute the Eulerian phase on CPUs and offload Lagrangian particle tracking to GPUs.

Several implementations that make use of such heterogeneous EL computations have been reported~\cite{xu2012,sweet2018,ge2020,dziekan2022,wang2022,liu2023,dennis2024,muralikrishnan2024}. All of them suffer from at least one of the following drawbacks, which restricts EL simulations to maximum particle numbers on the order of 1×10\textsuperscript{9}, and necessitating the development of more scalable approaches:
\begin{itemize}
  \item No use of heterogeneous architecture: Some of the implementations port both the calculation of the Eulerian and of the Lagrangian phase to the GPU, leaving the CPUs idle~\cite{liu2023,muralikrishnan2024}. 
  \item No two-way coupling: Some GPU-based EL algorithms only implement one-way coupling~\cite{sweet2018,wang2022}. While this assumption suffices for many scenarios, many disperse two-phase flows are two-way coupled. Examples include clouds~\cite{chen2025}, pneumatic conveying~\cite{lain2013}, or spray combustion~\cite{ruger2000}. Implementing two-way coupling complicates communications and increases memory demands.
  \item Synchronization barriers: Codes that make use of CPUs and GPUs at the same time usually encounter idling times in either the Eulerian or the Lagrangian part, as the calculation of a phase can only start when the source terms coming from the other phase are calculated. Some codes at least calculate both phases simultaneously on CPUs and GPUs~\cite{dziekan2022}, others do not~\cite{xu2012,sweet2018,wang2022}.
  \item No or complicated load balancing~\cite{xu2012,sweet2018,ge2020,dziekan2022,liu2023,dennis2024,muralikrishnan2024}: As the particles move freely through the domain, load balancing can become problematic. Particles may accumulate in small regions of the computational domain. With the traditional approach of dividing the domain into fixed partitions, this can cause huge load imbalances. CPUs/GPUs that have to handle many particles will take longer for their calculations, while those hosting fewer particles may idle a large fraction of the computation time. There are several approaches to tackle this problem, a brief overview is given in subsection~\ref{sec:decomposition}.
  \item Communication overhead: Employing a multitude of CPUs and GPUs requires communication between the devices, e.g.~for sending and receiving Eulerian field data or transferring particles between devices. This negatively influences scaling, especially when it comes to exascale-sized problems incorporating thousands of devices.
\end{itemize}

We propose SCALE-TRACK, a scalable two-way coupled Euler-Lagrange particle tracking algorithm, that enables high-fidelity EL simulations on exascale HPC systems. This algorithm overcomes the drawbacks listed above, or at least reduces their impact on scalability. It was developed as a sub-project of Inno4Scale ("Innovative Algorithms for Applications on European Exascale Supercomputers", funded by the EuroHPC Joint Undertaking). Inno4scale supported the development of advanced algorithms and applications for exascale HPC systems up to technology readiness level (TRL) 3, i.e.~the development of experimental proofs of concepts.

The Eulerian part in the simulations is calculated by third-party software such as OpenFOAM, which is used in the current implementation. However, the SCALE-TRACK architecture allows low-effort coupling to any CFD solver for the Eulerian phase.

SCALE-TRACK itself focuses on the Lagrangian part of EL simulations, and makes use of novel approaches for asynchronous EL coupling, chunk based partitioning, balanced initialization, cache-friendly data structures, and non-blocking communication strategies. Especially the novel asynchronous coupling algorithm and the parallelization strategies used distinguish it from other GPU based Lagrangian simulation software. We use the term asynchronous coupling to describe our code's core feature that the Eulerian and Lagrangian calculations can start without waiting for their counterparts to finish, so they are not necessarily calculating the same time steps at the same time. Ensuring that the communication between these two parts works properly, for example that sources are accumulated and transferred in a correct manner, has proven to be quite complex.

In section~\ref{sec:algo}, we describe the algorithm. Its correctness is demonstrated in section~\ref{sec:vali} by comparisons of test cases to an analytical solution and to simulations using the open-source flow simulation tool OpenFOAM (OF). Section~\ref{sec:scaling} shows scaling results from runs of another OF-based test case on the HPC cluster "MareNostrum5"~\cite{banchelli2025}. A summary and possible future developments are presented in section~\ref{sec:end}.

\section{\label{sec:algo}Description of the SCALE-TRACK algorithm}
In the following subsections we first describe the governing equations for the test cases presented later on. We then describe SCALE-TRACK's novel methods, and conclude with some implementation details.

\subsection{\label{subsec:govEq}Governing equations}
We developed two solvers in the course of the project. The first one features an incompressible, isothermal flow, the second a compressible, non-isothermal flow including buoyancy effects. In this first case, the Eulerian domain is governed by the continuity equation
\begin{equation}
  \nabla \cdot \mathbf{u}_{\mathrm{f}} = 0
\end{equation}
and the momentum equation
\begin{equation}
\frac{\partial \mathbf{u}_{\mathrm{f}}}{\partial t}
+ \nabla \cdot \left(\mathbf{u}_{\mathrm{f}} \mathbf{u}_{\mathrm{f}} \right)
- \nabla \cdot \left( \nu_{\mathrm{f}} \nabla \mathbf{u}_{\mathrm{f}} \right)
=
- \frac{1}{\rho_{\mathrm{f}}}\nabla p
+ \frac{\mathbf{S}_{\mathrm{u}}}{\rho_{\mathrm{f}}}.
\label{eq:momentum}
\end{equation}
in which $\mathbf{u}_{\mathrm{f}}$, $\rho_{\mathrm{f}}$, and $\nu_{\mathrm{f}}$ are the fluid's velocity, density, and kinematic viscosity, respectively. Further, $p$ is the pressure, and $\mathbf{S}_{\mathrm{u}}$ is the momentum source from the Lagrangian phase.

In the second case, the governing equations are somewhat more extensive. The continuity and momentum conservation equations are
\begin{equation}
\frac{\partial \rho_{\mathrm{f}}}{\partial t} + \nabla \cdot \left( \rho_{\mathrm{f}} \mathbf{u}_{\mathrm{f}} \right) = 0
\end{equation}
and
\begin{equation}
{\frac{\partial (\rho \mathbf{u}_{\mathrm{f}})}{\partial t}} + \nabla \cdot (\rho_{\mathrm{f}} \mathbf{u}_{\mathrm{f}} \mathbf{u}_{\mathrm{f}})
= -\nabla p + \rho_{\mathrm{f}} \mathbf{g}
+ \nabla \cdot \left( \mu_{\mathrm{f},\mathrm{eff}} (\nabla \mathbf{u}_{\mathrm{f}} + (\nabla \mathbf{u}_{\mathrm{f}})^T ) \right) + \mathbf{S}_{\mathrm{u}},
\end{equation}
with $\mathbf{g}$ being the gravitational acceleration, and $\mu_{\mathrm{f},\mathrm{eff}}$ being the effective dynamic viscosity of the fluid, comprising molecular and turbulent effects. Contributions from a moving reference frame are neglected.

An equation for the sensible internal energy $e_{\mathrm{f}}$ is also solved:
\begin{equation} 
\frac{\partial (\rho_{\mathrm{f}} e_{\mathrm{f}})}{\partial t}
+ \nabla \cdot (\rho_{\mathrm{f}} \mathbf{u}_{\mathrm{f}} e_{\mathrm{f}})
- \frac{\partial (\rho_{\mathrm{f}} K_{\mathrm{f}})}{\partial t}
- \nabla \cdot (\rho_{\mathrm{f}} \mathbf{u}_{\mathrm{f}} K_{\mathrm{f}}) 
+ \nabla \cdot (p \mathbf{u}_{\mathrm{f}})
= \nabla \cdot (\alpha_{\mathrm{f},\mathrm{eff}} \nabla e_{\mathrm{f}}) + \rho_{\mathrm{f}} \mathbf{u}_{\mathrm{f}} \cdot \mathbf{g} + S_{\mathrm{e}},
\end{equation}
where $K_{\mathrm{f}}$ is the fluid's kinetic energy per unit mass, $\alpha_{\mathrm{f},\mathrm{eff}}$ is its effective thermal diffusivity, including laminar and turbulent contributions, and $S_{\mathrm{e}}$ is the energy source from the Lagrangian two-way coupling.

Finally, a transport equation for water vapor $\rho_{\mathrm{v}}$ is necessary. Assuming only small gradients in $\rho_{\mathrm{f}}$, this is given by:
\begin{equation}
  \frac{\partial (\rho_{\mathrm{v}})}{\partial t} + \nabla \cdot (\mathbf{u}_{\mathrm{f}} \rho_{\mathrm{v}}) = \nabla \cdot (D_{\mathrm{f},\mathrm{eff}} \nabla \rho_{\mathrm{v}}) + S_{\rho_{\mathrm{v}}}.
\end{equation}
$D_{\mathrm{f},\mathrm{eff}}$ is the water vapor's effective diffusivity, again including molecular and turbulent contributions, and $S_{\rho_{\mathrm{V}}}$ is the Lagrangian water vapor source.

The water vapor mass transfer between the continuous (or gas) phase and the disperse (or liquid phase) is calculated as
\begin{equation}
  \frac{d m_{\mathrm{p}}}{dt} = 2 \pi D_{\mathrm{v}} d_{\mathrm{p}} \rho_{\mathrm{v},\mathrm{sat}} (S_{\mathrm{v},\mathrm{f}} - S_{\mathrm{v},\mathrm{p}}),
\end{equation}
where $D_{\mathrm{v}}$ is the diffusivity of water vapour, $\rho_{\mathrm{v},\mathrm{sat}}$ the water vapor density at saturation, and $S_{\mathrm{v},\mathrm{f}}$ and $S_{\mathrm{v},\mathrm{p}}$ the saturation wrt.~water vapor in the fluid and at the particle surface, respectively. The quantity $dt$ is the time step, $d_{\mathrm{p}}$ the particle diameter, and $m_{\mathrm{p}}$ the particle mass. The source term for the Eulerian field is
\begin{equation}
  S_{\mathrm{\rho_V}} = \frac{1}{V_{\mathrm{cell}}} \frac{d m_{\mathrm{p}}}{dt},
\end{equation}
where $V_{\mathrm{cell}}$ is the volume of the Eulerian grid cell containing the particle.

The particles' movement is governed by Newton's second law of motion:
\begin{eqnarray}
  \frac{d \mathbf{x}_{\mathrm{p}}}{dt} = \mathbf{u}_{\mathrm{p}}, \\
  m_{\mathrm{p}} \frac{d \mathbf{u}_{\mathrm{p}}}{dt} = \sum_{i} \mathbf{F}_{\mathrm{p},i},
\end{eqnarray}
where $\mathbf{x}_{\mathrm{p}}$ is the particle's position, $\mathbf{u}_{\mathrm{p}}$ its velocity, and $\mathbf{F}_{\mathrm{p},i}$ are the forces acting on the particle. For simplicity, only the drag force is considered here, using an empirical drag coefficient~\cite{schiller1933}. The source term for the Eulerian phase is calculated as
\begin{equation}
\label{eq:momentumSource}
  \mathbf{S}_{\mathrm{u}} = \frac{1}{V_{\mathrm{cell}}} \frac{d(m_{\mathrm{p}} \mathbf{u}_{\mathrm{p}})}{dt}.
\end{equation}

The thermal energy transfer from or to the Lagrangian phase is characterized by
\begin{equation}
  m_{\mathrm{p}} C_{\mathrm{p},\mathrm{p}} \frac{d T_{\mathrm{p}}}{dt} = \pi {Nu}_{\mathrm{p}} \kappa_{\mathrm{f}} d_{\mathrm{p}} (T_{\mathrm{f}} - T_{\mathrm{p}}) - L \frac{d m_{\mathrm{p}}}{dt},
\end{equation}
and the source term for the Eulerian phase is
\begin{equation}
  S_{\mathrm{e}} = \frac{1}{V_{\mathrm{cell}}} C_{\mathrm{p},\mathrm{p}} \frac{d (m_{\mathrm{p}} T_{\mathrm{p}})}{dt}.
\end{equation}
$C_{\mathrm{p},\mathrm{p}}$, $T_{\mathrm{p}}$, and $Nu_{\mathrm{p}}$ are the specific heat capacity of the particle, its temperature, and its Nusselt number. Quantity $\kappa_{\mathrm{f}}$ and $T_{\mathrm{f}}$ are the fluid's thermal conductivity and temperature, respectively, at the particle position. $L$ is the latent heat of vaporization of water.

\subsection{Domain decomposition and data structures}
\label{sec:decomposition}

\begin{figure}
\centering
\includegraphics[width=\linewidth]{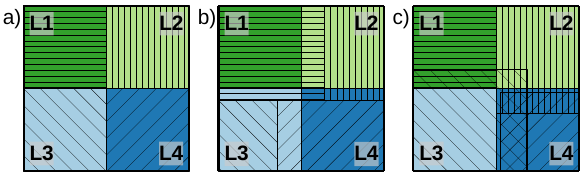}
\caption{\label{fig:decom} Domain decomposition strategies, showing Eulerian partitions via colors and Lagrangian partitions L1 to L4 via different hatching styles. a) Eulerian and Lagrangian domains are identical (the conventional and most common spatial partitioning} approach). b) Lagrangian partitions do not necessarily coincide with Eulerian ones. Additionally, they may move, grow and shrink. c) The same as in b), but Lagrangian partitions can overlap.
\end{figure}

There are two common approaches in EL simulations for distributing the computational domain to the processing devices. The most common method, spatial partitioning, is to split the fluid domain into several partitions and to use the same geometric decomposition for the particle phase, as shown in Fig.~\ref{fig:decom}a). This keeps the communication between devices for the Eulerian and the Lagrangian phase simple, as the exchange of Eulerian field data and particle source data only happens between the two coinciding partitions, e.g.~L1 only has to communicate with the dark green Eulerian partition. However, particles must be transferred between devices as soon as they leave their respective partition, which can happen multiple times for the same particle in a single Eulerian time step, thus increasing communication effort. Also, these transfers are not easily predictable, which makes scheduling harder. A further drawback is the possibility of load imbalances due to particle accumulation and dilution in the partitions, which leads to higher computational load in partitions with a higher particle count. Usually, the solution procedure cannot evolve until all partitions are done solving the current time step. Thus, load imbalances have a negative impact on total computational time.

Another, but less common approach, as depicted in Fig.~\ref{fig:decom}b), is to keep the Eulerian domain as before, but to partition the particles independently~\cite{muralikrishnan2024, sweet2018}. This improves the load balancing behavior, as the Lagrangian domains can also be readjusted independently of the Eulerian partitions by moving, growing, and shrinking. However, the Lagrangian partitions may now need information from multiple Eulerian partitions, e.g.~L1 needs information from all four of the Eulerian partitions. Usually, all the Eulerian data are sent to all devices, drastically increasing memory demands on the GPUs and also communication between CPUs and GPUs. Furthermore, particle transfer between the partitions is still required, which now may even increase as particles may have to swap partitions when these partitions grow and shrink.

A third approach, the particle sharing concept, is to freely distribute the Lagrangian particles among the computing devices. This comes with the drawback that the whole Eulerian fields are needed on all devices doing Lagrangian computations. To keep the communication at least local on computational nodes, this approach may also be coupled with hierarchical decomposition~\cite{thari2022}.

Weighted load balancing can be used for both of these approaches~\cite{golshan2023,stock2023}. Here, particles and/or cells get assigned a weight, depending on their numerical load. An algorithm then distributes cells and particles so that the summed loads of the different partitions are approximately equal.

For SCALE-TRACK, we use the approach shown in Fig.~\ref{fig:decom}c), allowing partitions to overlap: For instance, L4 overlaps with L2 and L3. When particles cross the borders of their current partition, the partition grows instead of transferring particles to a neighboring domain. Using this approach, the number of particle transfers between Lagrangian partitions can be drastically reduced, and such communications become more predictable.

In SCALE-TRACK, particles are represented by one or more chunks. Each chunk corresponds to a single Lagrangian partition, as illustrated in Fig.~\ref{fig:decom}c). A chunk is organized as structure of arrays, which positively impacts performance, especially on GPUs, by facilitating vectorization and coalesced memory access. Each GPU can handle one or more chunks. Handling multiple chunks on a single GPU can improve performance by "hiding" communication: After the GPU finishes computing one chunk, it can send the sources from the finished chunk to the CPU while already calculating the next chunk.

Rectangular bounding boxes are generated around the particles in each chunk. One purpose of these chunk bounding boxes is to identify the Eulerian partitions that are required for the calculation of a chunk, specifically only those that have an overlap with the respective bounding box. To reduce the size of the chunk bounding boxes during the initial and early stages of the simulation, an initialization procedure based on a Hilbert space-filling curve was implemented in SCALE-TRACK. The computational domain is discretized into a regular lattice of Hilbert voxels traversed by the Hilbert curve, which is subsequently partitioned into consecutive segments corresponding to the desired number of chunks. This approach provides flexibility in choosing the number of chunks while remaining highly scalable. Since the Hilbert index of each voxel is obtained through a deterministic global mapping, particles can be initialized independently on every process without any inter-process communication or synchronization. This property is particularly advantageous for the very large particle populations targeted by SCALE-TRACK, exceeding $2.56 \times 10^{11}$ particles.

Particles are randomly initialized within the Hilbert voxels belonging to their assigned segment, producing an approximately uniform particle distribution throughout the domain. The number of particles initialized in each Hilbert voxel is adjusted such that the chunks are completely filled with the prescribed number of particles, compensating for the fact that the number of voxels in each segment does not generally permit an identical number of particles per voxel. Since consecutive Hilbert segments remain spatially localized, the resulting chunks are compact and possess small bounding boxes during the initial and early stages of the simulation, thereby minimizing their overlap with Eulerian partitions and reducing the communication volume. As the simulation progresses, the bounding boxes change as the particles move. Dynamic chunk redistribution has been demonstrated as a proof of concept but is not part of the current production implementation.

During computation, having a compact organization of the chunks allows sending Eulerian data only from the few partitions that are actually required, e.g.~L1 in Fig.~\ref{fig:decom}c) only needs data from the upper-left Eulerian partition, and L2 only needs the two Eulerian partitions at the right.

\subsection{\label{subsec:algo} Asynchronous two-way coupling}

\begin{figure*}
\includegraphics[width=\textwidth]{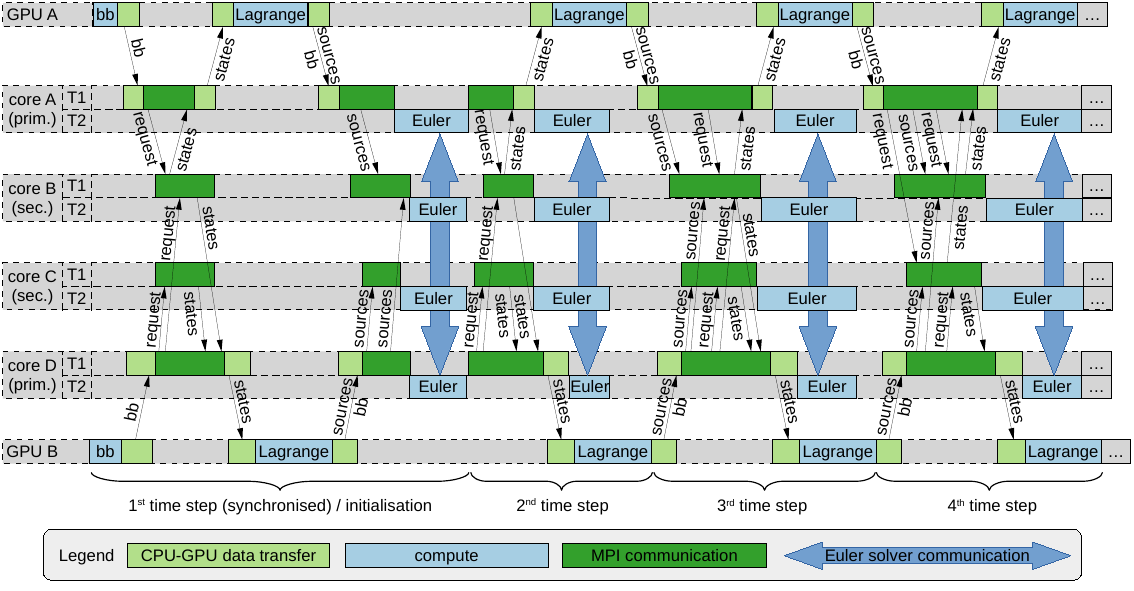}
\caption{\label{fig:timeline} A schematic of SCALE-TRACK's execution timeline with an exemplary setup comprising two GPUs A and B, two primary CPU cores (A and D), and two secondary CPU cores (B and C). Blue blocks are the actual computation routines, light green blocks are transfers between CPUs and GPUs, dark green blocks stand for MPI communication, blue arrows are OpenFOAM communications. Block sizes do not correspond to computational time. (T: thread; bb: bounding box).}
\end{figure*}

SCALE-TRACK's core feature is its asynchronous two-way coupling, which enables the simultaneous processing of Eulerian and Lagrangian parts on CPUs and GPUs, respectively. As explained in section~\ref{sec:intro}, SCALE-TRACK does not implement modeling of the Eulerian phase but uses an external software for this task, which is OpenFOAM in the example provided below. 

A schematic execution timeline with two GPUs, two primary and two secondary CPU cores is given in Fig.~\ref{fig:timeline}. The number of CPUs and GPUs is not limited by the algorithm apart from the fact that a primary CPU core is able to drive only a single GPU. Such cores are also called host CPU cores as shown in Fig.~\ref{fig:timeline}: GPU A's host is core A, GPU B's host is core D. This is evident in the flowchart, in which the light green boxes, which identify CPU-GPU data transfers, appear exclusively for the primary cores and the GPUs. The transfer and communication paths are identified by thin black arrows, pointing from the source to the target of the message. The transferred data are provided above the arrows. In the following, we explain the execution timeline in more detail.

We first consider the GPU workloads. At initialization, the particles are created directly on the GPUs, the corresponding bounding boxes are calculated, and transferred to the respective primary CPUs. As soon as the GPUs receive the required states (such as Eulerian velocity fields), the particle tracking is processed. After finishing the calculation of a Lagrangian time step (blue boxes), the GPUs send the updated bounding boxes (bb) of their chunks and the calculated sources, e.g.~momentum source $\mathbf{S_u}$ from Eq.~\ref{eq:momentumSource}, to their host cores. After receiving the states from the next Eulerian time step, the GPUs start calculating the next Lagrangian time step and the process repeats.

The CPU timeline is more complex. Each CPU core runs two threads, T1 and T2. T1 takes care of communication and scheduling, T2 performs the Eulerian computation, whereby the threads coordinate execution and resource management using locks and wait/notify mechanisms. Within the host cores, T1 manages CPU-GPU data movement and acts as the GPU’s controlling driver.

During the initialization phase, the Lagrangian and Eulerian computations are synchronized, meaning that the latter starts only when the data from the former become available. First, the host cores A and D obtain the bounding boxes from the GPUs and, via MPI (Message Passing Interface, dark green boxes), request the states of the Eulerian fields from other CPU cores that are needed to perform the Lagrangian calculation. The resulting sources are transferred from the GPUs to the hosts, which forward them to the appropriate secondary cores in order to perform the Eulerian computation. Since the latter is performed by an external solver, its communication is not detailed but noted by a blue double arrow. In the case of OpenFOAM, the Eulerian calculation is synchronized at the end or, in other words, it finishes at the same time for all partitions.

After the initialization, the following Eulerian and Lagrangian computations are asynchronous, i.e.~may be performed simultaneously. In order to accomplish this, the sources need to be extrapolated based on the available state. The details of this process are explained in section~\ref{subsec:ecm}. At the beginning of the second time step, the states are requested, communicated and transferred to GPUs, the sources are extrapolated and the Eulerian solver is executed in parallel to the Lagrangian one. For the following time steps the communication pattern repeats itself.

A closer look at the communicated sources and states reveals important algorithm features that are covered by a detailed description in the following. During the second and third time steps: 
\begin{itemize}
    \item GPU A requires field states from cores A and B. Core A requests the states from core B. Core B sends them to core A, and core A copies these and its own states to the GPU.
    \item GPU B needs fields from cores B, C, and D. Thus, its host core D requests and receives those fields, and copies them to GPU B. It also sends the sources to the respective cores.
\end{itemize}

At the fourth time step, the Lagrangian particles evolve such that GPU A additionally requires the Eulerian field states from core C, whereas the particle computation on GPU B no longer depends on the states from core B. These changes are reflected in the corresponding request and state communication patterns. Notably, the most recent source terms are communicated to the Eulerian partitions following the communication pattern of the previous time step to correct the extrapolated sources (see section~\ref{subsec:ecm} for details).

We think it is worth to mention that due to the asynchronous nature of the solver, the results are not fully deterministic and reproducible, as they depend on the initialization procedure, which might be different on different hardware setups, and on the number and type of involved devices and their execution speed. This might seem unfavorable, but in this case it is the cost for fully exploiting the resources of modern, heterogeneous exascale HPC systems. And as previously mentioned, conventional Euler-Lagrange algorithms are also not the true solution. So as long as the errors introduced by SCALE-TRACK are on the same order of magnitude as in conventional algorithms, we argue that the possible improvements outweigh this drawback.

\subsection{\label{subsec:ecm}Extrapolator-corrector method}

To minimize errors introduced by the asynchronous coupling, an extrapolator-corrector method for the Lagrangian sources is introduced. Since our approach is to keep the GPUs busy, we assume that usually the CPUs are ahead of the GPUs, i.e.~when starting to calculate the Eulerian phase, the Lagrangian sources for the current time step may not be available. For simplicity, we present the equations for a maximum delay of one time step.

The corrector for the current time step is the difference between the actual Lagrangian source term from the last time step, $S^{n-1}$, which is now available, and the estimated source from the last time step, $S^{n-1}_{\text{est}}$:
\begin{equation}
 {\Delta S^n_{\text{corr}} = S^{n-1} - S^{n-1}_{\text{est}}}
\end{equation}
By adding an extrapolator for the source during the current time step, $S^n_{\text{ext}}$, the estimated source for the current time step reads as
\begin{equation}
 {S^n_{\text{est}} = \Delta S^n_{\text{corr}} + S^n_{\text{ext}} = S^{n-1} - S^{n-1}_{\text{est}} + S^n_{\text{ext}}}.
\end{equation}
Since the estimated source from the previous time step is subtracted in the current time step, the procedure is conservative over time. If the sources become available with a delay of more than one time step, the corrector is the sum of all unincorporated sources from previous time steps minus the sum of the estimated sources from these time steps.

The extrapolators we investigated are:
\begin{subequations}
\begin{equation}
 {S^n_{\text{ext, zero}} = 0},
\end{equation}
\begin{equation}
 {S^n_{\text{ext, const}} = S^{n-1}},
\end{equation}
\begin{equation}
 {S^n_{\text{ext, lin}} = 2S^{n-1} - S^{n-2}},
\end{equation}
\end{subequations}
The first one, $S^n_{\text{ext, zero}}$, will be referred to as "zero extrapolator" and simply sets the extrapolated value to zero. If no source is available, then no source is added to the Eulerian fields. As soon as previous sources become available, they are incorporated. $S^n_{\text{ext, const}}$, the "constant extrapolator", sets the estimate to the true value of the source from the last available time step. The linear extrapolator, $S^n_{\text{ext, lin}}$, as the name suggests, linearly extrapolates from the values of the true sources of the last two known time steps.

The equations above only hold in the case of a constant time step. If the time increment varies between time steps, this must be accounted for either by scaling the source terms with the time-step ratio or by providing them as a rate of change. In subsection~\ref{subsec:analytical}, the three correctors described here are analyzed with respect to the errors they introduce.

\subsection{Further implementation details}
SCALE-TRACK is written entirely in the Julia programming language~\cite{bezanson2017} (version 1.10.8). It makes extensive use of concurrent and parallel tasks, using co-routines and threads. Execution of the Lagrangian kernel is in principle possible on CPUs as well as on GPUs. The CPU implementation was used mainly for testing and debugging and is not parallelized.

SCALE-TRACK is agnostic to the continuous phase solver. It requires information about the domain geometry, the mesh, and the necessary Eulerian fields. SCALE-TRACK feeds back the source terms from the disperse phase to the Eulerian fields.

For demonstration purposes, we coupled SCALE-TRACK to OF v2406~\cite{of2406}. The underlying solvers, "icoFoam" and "buoyantPimpleFoam", are extended by including additional Eulerian scalar fields (see subsection~\ref{subsec:govEq}). These additional fields are the sources from the Lagrangian phase and, in the case of buoyantPimpleFoam, water vapor concentration, since the test cases include the simulation of a cloud chamber. Here, OF is the governing application, which calls Julia functions when necessary.

The interpolation of Eulerian field values to a particle is performed using a nearest-grid-point scheme, wherein each particle adopts the field value of the computational cell it is currently in. Conversely, source terms contributed by a particle are deposited exclusively into the cell containing the particle.

The semi-implicit Euler method is employed for time integration of both, particle motion and thermal energy. For mass transfer calculations, an explicit Euler algorithm is used. Both methods demonstrate first-order convergence. The same methods are used in OpenFOAM.

In the current implementation, the two phases evolve asynchronously so that their physical times differ at most by one time step. The algorithm ensures continuous GPU utilization whenever possible, with CPUs pausing if they advance too far ahead. As GPUs are usually the cost-driving factor, our goal is to utilize them to their full potential. With that aspect in mind, a higher gap in time steps would make sense only if the CPUs are working at full load and the GPUs would have idling times while waiting for the CPUs, a scenario we aim to avoid.

As described in section~\ref{sec:decomposition}, the partitioning between the Eulerian and Lagrangian frames differs. Moreover, the Lagrangian partitions change their spatial extents, leading to changing overlaps with the Eulerian partitions. This behavior requires special treatment of the communication of field states and sources, as the messaging partner ranks must be identified anew at each time step. We reduce the idling time of CPUs by using the non-blocking consensus~\cite{hoefler2010} to identify the correct communication partners.

\section{\label{sec:vali}Verification and benchmarking}
\subsection{\label{subsec:analytical}Verification} 
\begin{figure}
\centering
\includegraphics[width=0.5\linewidth]{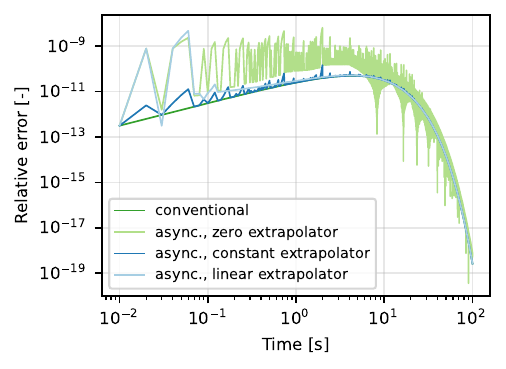}
\caption{\label{fig:instErrors} Evolution of the instantaneous relative error $e_{rel}$ for the continuous phase momentum at an exemplary time step of \SI{0.01}{s}. The error is calculated as the absolute difference between numerical and exact solutions at each time step in relation to the exact solution.}
\end{figure}

of the extrapolator-corrector method against an analytical solution
To test the extrapolator-corrector method that reduces the errors of the asynchronous two-way coupling, results for different extrapolators (see subsection~\ref{subsec:ecm}) are compared to an analytical solution~\cite{martinez2009}. A particle traveling at \SI{1}{\m \s\tothe{-1}} is placed within a quiescent fluid. The particle slows down, while the fluid absorbs its kinetic energy due to the two-way coupling and starts moving. Fig.~\ref{fig:instErrors} shows the relative errors of the momentum from the asynchronous approach using different extrapolation strategies and the conventional approach against the analytical one for the Eulerian phase at an exemplary time step of \SI{0.01}{s}. The relative error is calculated as
\begin{equation}
    e_\text{rel} = \left | {{\frac{p_\text{num} - p_\text{exact}}{p_\text{exact,total}}}} \right |,
\end{equation}
in which $p_\text{num}$ is the integral momentum in the respective phase of the numerical approach, $p_\text{exact}$ is the integral momentum in the same phase from the analytical solution, and $p_\text{exact,total}$ is the analytically calculated momentum of both phases.

The first solution in Fig.~\ref{fig:instErrors} is the conventional EL approach, i.e.~the Eulerian time step is solved, the Eulerian fields are sent to the particles, then the Lagrangian time step is solved, and the sources are sent to the Eulerian fields. It is worth noting that even this conventional method introduces errors, since processes that in reality happen simultaneously and continuously are treated subsequently and with discrete time steps.

The next solution strategy is to use the asynchronous algorithm with the zero extrapolator. This means that the Eulerian and the Lagrangian solvers use the sources or fields, respectively, that are available when they start solving for the current time step. Of course this method introduces comparatively large errors, sometimes several orders of magnitude large than in the conventional solution. It also introduces oscillations, as the sources can accumulate over several time steps, i.e the error accumulates. The error then suddenly drops when the sources become available and are fed back to the system.

For the next method, the Eulerian fields are corrected with the constant extrapolator. This means that the last available source is used as an extrapolator for the current source. This reduces the introduced error to the same level as the conventional method. The oscillations are also much weaker than with the zero extrapolator approach, as parts of the sources are incorporated at every time step. Eventually, the oscillations become negligible.

Finally, the linear extrapolator is applied. During the first time steps, large oscillations in the Eulerian phase arise, on a similar level as when using no extrapolated sources. After a few time steps, the oscillations vanish and the errors reduce to a level comparable to the conventional and the constant extrapolator approaches, but are still slightly larger.

\begin{figure}
\centering
\includegraphics[width=0.5\linewidth]{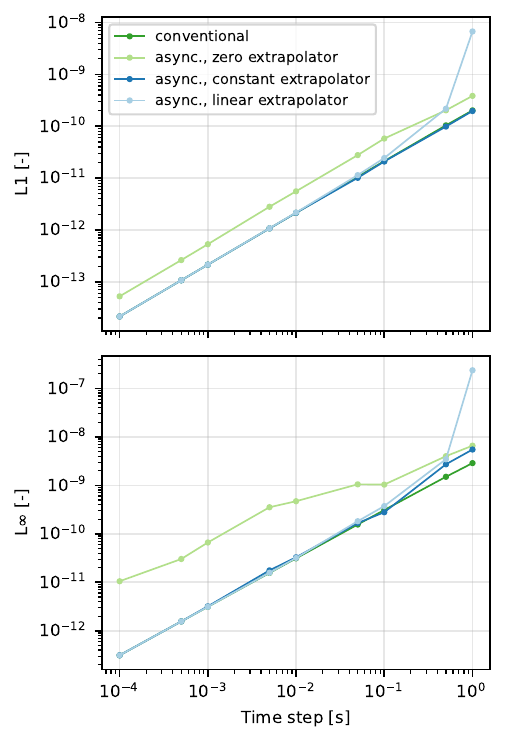}
\caption{\label{fig:analytical} Convergence behavior of different extrapolation approaches for the corrector. L1 and L$\infty$ error norms for the continuous phase are plotted versus time step size for the conventional method and three asynchronous extrapolation schemes (zero, constant, and linear extrapolators).}
\end{figure}

Fig.~\ref{fig:analytical} shows the error norms $L1$ and $L\infty$ for the conventional method and for the asynchronous calculations with the three different extrapolation approaches. The errors shown are of the continuous phase, those of the disperse phase behave similarly. Error calculations exclude the first 10\% of time steps to eliminate transient effects. All methods exhibit first-order convergence, at least for the $L1$ norm, due to the underlying first order solution algorithm.

The constant extrapolation method shows almost the same behavior as the conventional approach. Only for $L\infty$ at the largest time steps it reveals slightly larger error norms. The linear extrapolation behaves the worst at large time steps, but as the time step decreases, it also exhibits the same error level as the conventional method. Doing no extrapolation leads to higher errors, and for $L\infty$ it even reduces the order of convergence.

\subsection{\label{subsec:valOF}Benchmarking of SCALE-TRACK against results from OpenFOAM with a convection cloud chamber test case}

\begin{figure}
\centering
\includegraphics[height=0.3\textheight]{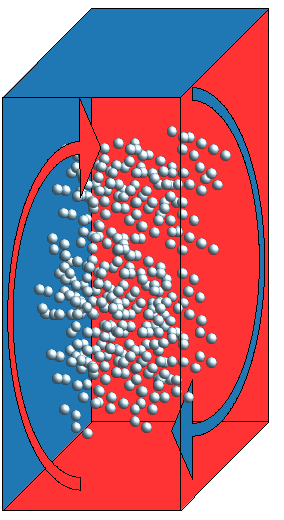}
\caption{\label{fig:chamberPrinciple} Schematic of a convection cloud chamber: a buoyancy-driven flow is induced by a combination of cool (blue, $T_{cool} = $ \SI{283}{\K}) and a warm (red, $T_{warm} = $ \SI{293}{\K}) boundaries. The walls at the front (cold) and at the right (warm) are removed for illustration purposes. All walls are water-saturated. Droplets, as indicated by the spheres (not to scale) move through the domain and grow and shrink depending on the humidity they encounter.
}
\end{figure}

A convection cloud chamber is employed as a test case for the comparison of OF's built in Lagrangian tracking with SCALE-TRACK. A modified version of "buoyantPimpleFoam" was used with governing equations for compressible, non-isothermal flow as described in subsection~\ref{subsec:govEq}. Cloud chambers are used in laboratory experiments for cloud research~\cite{shaw2025}. Some operate on the principle of a Rayleigh-Bénard cell~\cite{chen2025}, inducing a buoyancy-driven flow with a cool upper boundary and a warm lower boundary. When the aspect ratio (height to width) becomes too large for pure Rayleigh-Bénard convection, side walls can also be cooled or heated to enhance convection~\cite{wang2024} (see Fig.~\ref{fig:chamberPrinciple}). Additionally, the chamber walls are water saturated, thereby serving as a source for water vapor for the continuous phase. Mixing of cold and warm water-saturated air results in supersaturation, which is necessary for the formation and growth of cloud droplets.

Typically, cloud condensation nuclei (CCN) are continuously injected in this type of cloud chambers. The CCN form new droplets under supersaturated conditions, thereby balancing the loss of droplets at the boundaries. To keep the model simple, in this test case droplets do not vanish at wall contact, but are reflected, and no CCN are injected. Two-way coupling allows for the exchange of momentum, heat, and mass between the continuous and dispersed phase as described by the governing equations in subsection~\ref{subsec:govEq}.

The test case features a cloud chamber with a cross section of \SI{3}{\m} × \SI{3}{\m} and a height of \SI{9}{\m}, decomposed into 96×96×288 elements, resembling a cloud chamber large enough to form drizzle\cite{shaw2025,shaw2025design,wang2024}. Typical droplet number concentrations in such cloud chambers can reach \SI{1000}{\cm\tothe{-3}} or more, resulting in 81×10\textsuperscript{9} droplets. Since this would not be tractable on the available workstation (dual AMD EPYC 9684X CPU with a total of 192 cores, 756 GB of RAM, and an NVIDIA RTX 6000 Ada GPU), we make use of the parcel approach, combining identical particles to a so-called parcel~\cite{crowe1977}. To demonstrate what is possible on that workstation, we used 0.14×10\textsuperscript{9} parcels in OF, and 1.4×10\textsuperscript{9} parcels for SCALE-TRACK, using almost all of the CPU's memory in OF's case, and almost all of the GPU's memory for SCALE-TRACK. We chose 500 droplets per parcel for OF, and 50 droplets per parcel for SCALE-TRACK, resulting in a total of 70×10\textsuperscript{9} particles for both and a number concentration of approx.~\SI{864}{\cm\tothe{-3}}. The droplets or parcels are tracked through the domain for \SI{3}{\s} at a time step of \SI{5}{\ms}. Before initializing the parcels randomly in the domain, the flow fields were run in single phase mode until a statistically stable flow was achieved. The bottom and two adjacent side walls were set to $T_{warm} = $ \SI{293}{\K}, and the top and the remaining two walls were set to $T_{cold} = $ \SI{283}{\K} (see Fig.~\ref{fig:chamberPrinciple}). All walls are water-saturated.

\begin{figure}
\includegraphics[width=\linewidth]{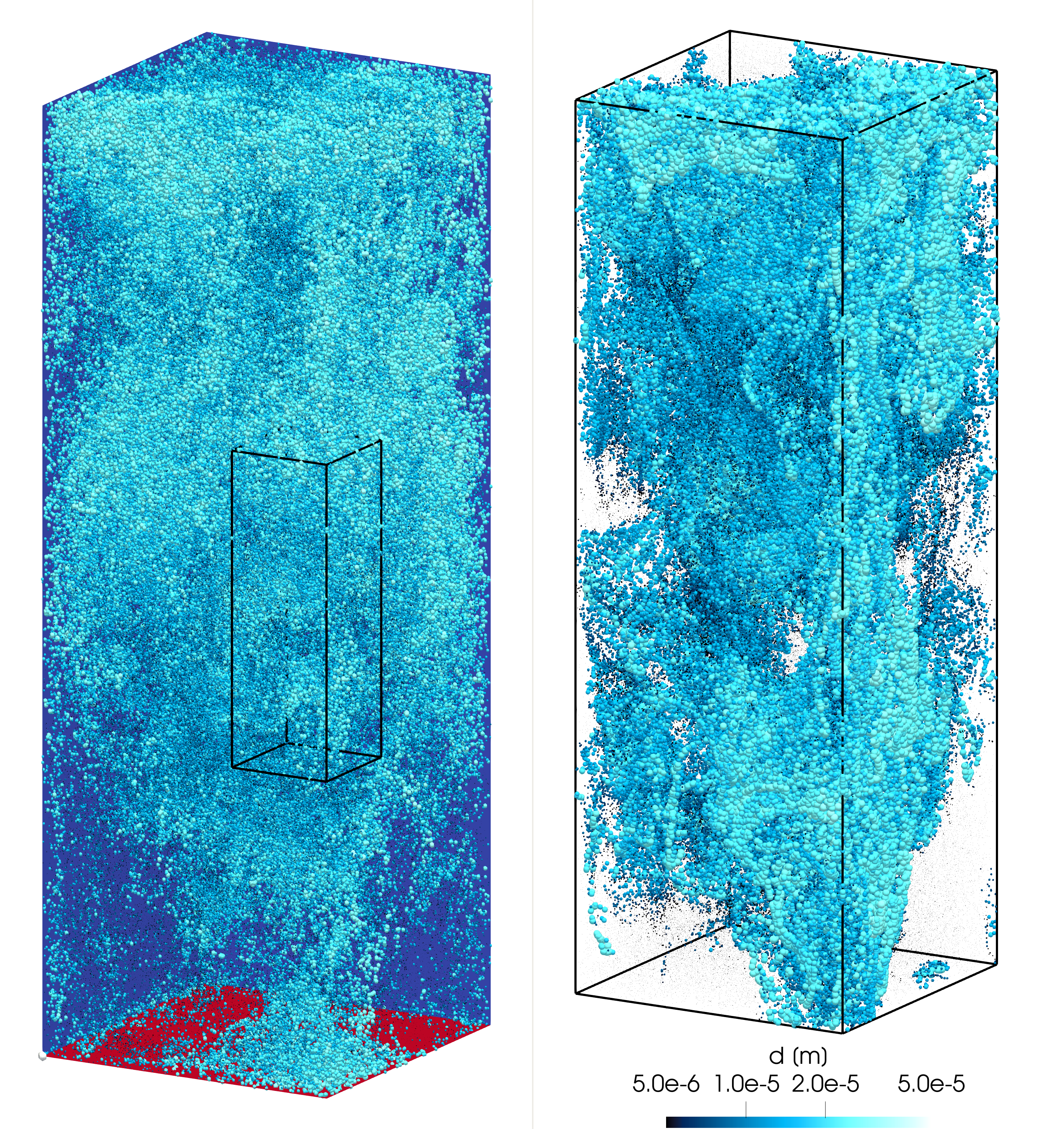}
\caption{\label{fig:chamberResult} 
Snapshots of a cloud chamber simulation. The whole simulation domain is shown on the left hand side, with the two warm side walls and the cold top wall removed for visibility, and a random selection of droplets. The black cuboid on the left hand side is zoomed in on the right hand side, showing a more detailed view of the droplets. The droplets are scaled according to their diameter, but magnified.
}
\end{figure}

\begin{figure}
\centering
\includegraphics[height=0.5\textheight]{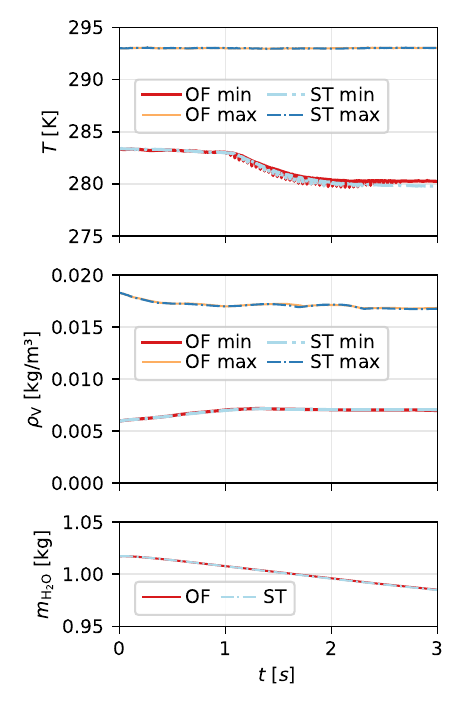}
\caption{\label{fig:chamberComparison} Comparison for cloud chamber simulations with OpenFOAM (OF) and SCALE-TRACK (ST) Lagrangian tracking. Top: minimum and maximum values of temperature $T$ in the domain over time. Middle: maximum and minimum water vapor concentration $\rho_\mathrm{V}$ in the domain over time. Bottom: Total water vapor mass $m_\mathrm{H_2O}$ in the domain over time.
}
\end{figure}

A snapshot of the simulation is shown in Fig.~\ref{fig:chamberResult}. The minimum and maximum values of temperature and water vapor content in the computational domain, along with the integral water vapor mass over the Eulerian domain are used to compare the two methods (Fig.~\ref{fig:chamberComparison}). Slight deviations in the temporal evolution of the minimum and maximum temperature values are observed. After \SI{1}{\s}, small oscillations occur in the minimum values for both solvers, with small deviations between the two thereafter. With respect to the temperature difference between the bottom and the top of the simulation domain, the maximum deviations are 0.79\% and 5.8\% for the maximum and minimum temperature, respectively. Almost no deviations are visible for the minimum and maximum water vapor content values. Scaled again with respect to the vapor density difference between bottom and top, the maximum deviations are 2.5\% and 1.3\% for maximum and minimum values. The domain integral of the water vapor mass also exhibits no visible deviations in the graph. Here, the maximum deviation is 0.48\% with respect to OpenFOAM's solution.

SCALE-TRACK processed ten times as many parcels as OF. Nevertheless, it achieved a 2.7-fold faster time-to-solution and a 2.5-fold better energy-to-solution.

\subsection{Discussion}
The comparison against the analytical solution reveals that an asynchronous two-way coupling EL approach can be used with relative errors at the same level as those of the conventional EL approach. However, it is essential to use a suitable extrapolator-corrector method for the source terms in the Eulerian field when employing an asynchronous coupling. A correction with zero extrapolation already reduces the introduced errors to possibly acceptable levels. However, this method may lead to oscillations in the Eulerian phase, as can be seen in Fig.~\ref{fig:instErrors}. These oscillations are caused by the build-up of unused sources and the sudden feeding of these back into the system when they become available, resulting in much higher source magnitudes than the average. Depending on the ratio between the sources and other forcings within the simulation, this may cause numerical instabilities. The constant extrapolation method, i.e.~using the true source values from the last available time step, further reduces the error to a level similar to the conventional treatment used in EL simulations. It also drastically reduces the oscillations, although some smaller ones are still discernible. The linear extrapolator introduces large errors and oscillations at the beginning and may also do so when large temporal gradients occur. Over time, the oscillations vanish, but this method seems to introduce a higher error than the constant extrapolator. As the computational effort for the constant extrapolator is lower than for the linear extrapolator and its ability to reduce the errors is the best among the tested methods, we suggest to use the algorithm with this correction.

Comparing OF and SCALE-TRACK reveals small deviations in the temporal evolution for the minimum temperature values across the entire domain. A possible explanation for the deviations is that OF uses a more sophisticated tracking method than SCALE-TRACK. SCALE-TRACK simply divides the Eulerian time step into a fixed number of sub-steps and then tracks the particles for the duration of the sub time step before updating the interpolated values from the Eulerian phase. OF, on the other hand, uses a concept similar to the Courant-Friedrichs-Lewy (CFL) condition~\cite{lewy1928}: Particles are tracked for a maximum distance of a factor (sensibly less than one) multiplied by a cell length scale, until the interpolation from the continuous phase is updated. OF also decomposes Eulerian cells into tetrahedra and updates the interpolated values as soon as a particle hits a face of a tetrahedron. Even slight deviations in the locations of the particles may lead to large differences later along their trajectories. Adding to this, the number of parcels tracked in SCALE-TRACK was ten times higher. The parcels are also initialized randomly across the domain. Both reasons may lead to differences between the two simulations. 

The simulation of 1.4×10\textsuperscript{9} droplets on a workstation with a single GPU is on the same order of magnitude as previous HPC simulations in terms of number of particles tracked, which highlights the efficient resource usage of SCALE-TRACK. We could have also run a SCALE-TRACK simulation with fewer particles, but the Eulerian domain was the driver in computational time, i.e.~the Lagrangian part on the GPU was calculated faster than the Eulerian part on the CPU. Thus, reducing the parcel number in SCALE-TRACK would have had no effect on run times.

Overall, we conclude that the proposed asynchronous tracking algorithm achieves an accuracy comparable to the conventional EL approach, while enabling higher-fidelity EL simulations on a local workstation in less time and with lower energy consumption.

\section{\label{sec:scaling}Scaling}

The cloud chamber from subsection~\ref{subsec:valOF} was reused for scaling runs of the SCALE-TRACK algorithm. The following scaling diagrams show, on one hand, timings for the calculation of the Eulerian part including all the communication from OF (compute Eulerian blocks in Fig.~\ref{fig:timeline}). Furthermore, the Lagrangian timings include only the computation of the particle tracking kernel (compute Lagrange block in Fig.~\ref{fig:timeline}) excluding the transfer and communication of the states and sources. However, the total timings represent the wall-clock time of the complete solver execution. It should be noted that the sum of the two compute timings is in most cases larger than the total timing, since the former are executed asynchronously.

\subsection{Strong scaling}

\begin{figure}
\centering
\includegraphics[width=\linewidth]{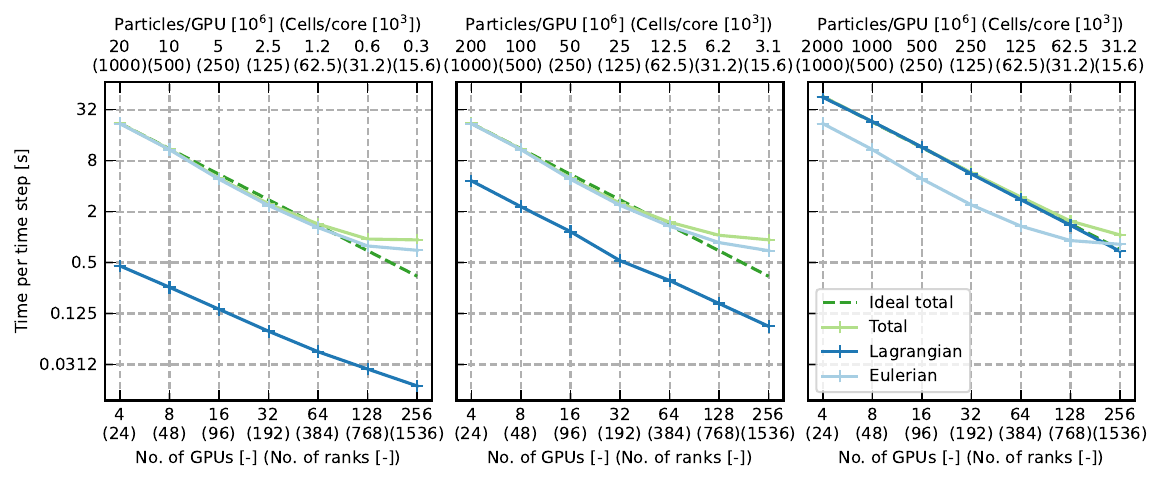}
\caption{\label{fig:strongScaling} Strong scaling results in terms of computational time per time step for a cloud chamber with 24×10\textsuperscript{6} Eulerian cells and, from left to right, 8×10\textsuperscript{7}, 8×10\textsuperscript{8}, and} 8×10\textsuperscript{9} Lagrangian particles. Shown are times for Eulerian and Lagrangian parts of the algorithm, the total time per time step, and a reference for ideal scaling.
\end{figure}

Strong scaling runs, i.e.~increasing the computational resources (number of CPUs and GPUs) while keeping the problem size (number of cells and parcels) constant, were performed on LUMI's GPU partition, where each node consists of a 64-core AMD EPYC 7A53 CPU, 512 GB of RAM, four AMD MI250X GPUs with two compute dies each and a total of 128 GB of HBM2 memory. Results are shown in Fig.~\ref{fig:strongScaling}. The Eulerian grid was composed of 24×10\textsuperscript{6} cells (200x200x600), while the number of particles was fixed at 8×10\textsuperscript{7}, 8×10\textsuperscript{8}, and 8×10\textsuperscript{9}, respectively. Each MPI rank was assigned two cores to improve performance, reflecting the two-threads-per-process approach described in subsection~\ref{subsec:algo}.

As expected, higher particle numbers lead to higher computational times per time step. For the two smaller test cases, the Eulerian part takes more time. For the largest case the Lagrangian computation usually takes more time, except for the configuration with the highest node count. Regardless of which part takes longer, the total time per time step is mostly approximately the maximum of the Lagrangian and Eulerian computation times, showing that running them concurrently works well.

The Lagrangian component demonstrates ideal scaling at 8×10\textsuperscript{9} particles. Even at the relatively low load of 31.2×10\textsuperscript{6} particles per GPU, kernel execution overhead remains minimal despite underutilized GPU resources, achieving a parallel efficiency of 1.03. At 8×10\textsuperscript{8} particles however, scaling begins to deviate from ideal behavior at 64 GPUs and 12.5×10\textsuperscript{6} with a parallel efficiency of 0.94, dropping further to 0.82 at 256 GPUs. This aligns with the observation from the larger case, where the minimum particle count per GPU was 31.2×10\textsuperscript{6} and parallel efficiency was still excellent. That is approximately the same load as in this case with 32 GPUs. For 8×10\textsuperscript{7} particles, the configuration with the fewest nodes yields only 20×10\textsuperscript{6} particles per GPU, which is already below the minimum observed in the largest case and falls within the region where the medium-sized case begins to deviate from ideal scaling. Consequently, parallel efficiency is already low with 0.89 for 8 GPUs and declines further to 0.42 for 256 GPUs.

The Eulerian component begins to deviate from ideal scaling at 768 ranks (31,200 cells per core), adversely affecting the total time per time step. For the largest test case, this means that the computation time for the Eulerian phase becomes larger than that for the Lagrangian phase, adversely influencing the total computation time.

Beyond this point, the overhead from transfer and communication in the Euler-Lagrange coupling becomes significant. This is most apparent at 1536 ranks for the 8×10\textsuperscript{9} particle case, where the Eulerian and Lagrangian computations take approximately equal time. Since these calculations run concurrently, the difference between their maximum and the total time represents the Euler-Lagrange coupling overhead.

\subsection{Weak scaling}

\begin{figure}
\centering
\includegraphics[width=0.5\linewidth]{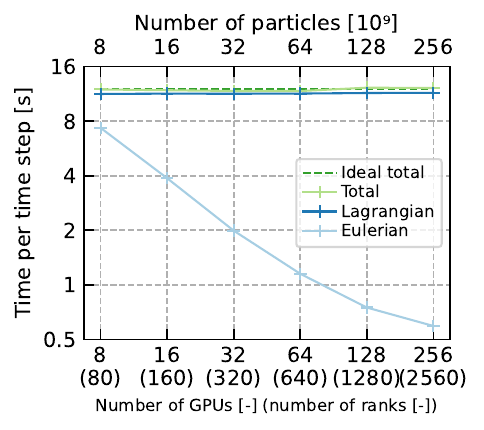}
\caption{\label{fig:weakScaling} "Semi"-weak scaling results in terms of computational time per time step for a cloud chamber with 24×10\textsuperscript{6} Eulerian cells (held constant) and 1×10\textsuperscript{9} Lagrangian particles per GPU, resulting in up to 256×10\textsuperscript{9} particles, tracked on 256 GPUs. Shown are times for Lagrangian and Eulerian parts of the algorithm, the total time per time step, and a reference for ideal scaling.}
\end{figure}

"Semi-weak" scaling runs were performed on MareNostrum5's accelerated partition~\cite{banchelli2025}. Each node features a dual Intel Sapphire Rapids 8480+ with 80 cores in total, 512 GB of RAM, and four NVIDIA Hopper H100 GPUs, each with 64 GB of memory. They are termed "semi"-weak because the Eulerian problem size was fixed at 24×10\textsuperscript{6} cells, and only the Lagrangian part was actually used for weak scaling, i.e.~keeping the problem size (number of particles) per resource (GPU) constant. We used 4×10\textsuperscript{9} particles per node or 1×10\textsuperscript{9} particles per GPU. Fig.~\ref{fig:weakScaling} shows the results.

As previously, the computational times per time step are shown separately for Eulerian and Lagrangian parts, along with a reference for ideal scaling. SCALE-TRACK achieves nearly ideal weak scaling for the Lagrangian part on up to 256 GPUs, with the largest run totaling 256×10\textsuperscript{9} particles. The total time also appears to scale nearly ideal. The Eulerian part exhibits the same behavior as in the strong scaling runs and starts to deviate from ideal scaling at 640 ranks, equaling 37,500 cells per core.

\subsection{Discussion}
The results indicate that the SCALE-TRACK algorithm performs as intended, showing excellent strong and weak scaling behavior. In the strong scaling experiments, the total computational time begins to deviate noticeably from ideal scaling at 768 MPI ranks. This deviation can be mainly attributed to the Eulerian component of the simulation, which also departs from ideal scaling at the same rank count, specifically when the number of cells per core decreases below approximately 40,000. Such behavior is well documented in Eulerian simulations~\cite{galeazzo2024}, in which communication overhead becomes increasingly dominant as the computational workload per process decreases. However, the EL coupling has its own overhead in the form of the communication of field states and sources, which becomes more pronounced for a higher number of ranks. This is to be expected and should be mitigated by ensuring that the Eulerian computation is short enough such that this communication can be performed while the Lagrangian part is still being calculated.

Furthermore, Fig.~\ref{fig:strongScaling} demonstrates that the Eulerian and Lagrangian components operate concurrently with high efficiency, provided that the Eulerian workload remains sufficiently large. Up to 384 ranks, the total time per time step is substantially lower than the sum of the individual execution times of both components, indicating effective parallel overlap. An approximate lower bound for efficient parallelization has also been identified at approximately 1\% of the maximum particle load the GPUs can handle.

The semi-weak scaling results highlight SCALE-TRACK's capability to efficiently exploit heterogeneous exascale HPC systems: Up to 256×10\textsuperscript{9} particles have been tracked with near-ideal scaling. It should be possible to use even larger numbers of nodes and GPUs without deviating from ideal scaling. Notably, tracking 256×10\textsuperscript{9} particles surpasses previously reported studies known to the authors by approximately two orders of magnitude in terms of particle count.

\section{\label{sec:end}Summary and outlook}
With SCALE-TRACK we developed a novel approach for two-way coupled Euler-Lagrange simulations on heterogeneous computing architectures, in which the Eulerian and Lagrangian phases are calculated asynchronously. It is a successful proof-of-concept for our novel approaches that ensure to make proper use of modern heterogeneous exascale HPC systems, as this was the goal of the Inno4Scale project that funded the research.

The asynchronous coupling algorithm has been explained in detail with the aid of a schematic work flow for an exemplary hardware setup. Furthermore, SCALE-TRACK introduces a new approach to decompose the Lagrangian phase independently of the Eulerian, and reduces communication overhead through cache-friendly data structures. This novel idea has also been explained in depth, discussing its principal ideas and its advantages and disadvantages.

We have successfully verified SCALE-TRACK's accuracy using an analytical solution and by benchmarking it with the built-in capabilities of OpenFOAM. The extrapolator-corrector method we proposed successfully reduces the errors introduced by the asynchronous calculation to a level as low as that of conventional EL simulations.

Scaling runs revealed the excellent scaling of the presented algorithm. A maximum number of 256×10\textsuperscript{9} tracked particles, distributed over 64 nodes with 256 GPUs, was achieved. This marks a significant increase over previous simulations and may be a base for future applications in high-fidelity EL simulations. Strong scaling also showed excellent results down to utilizing the GPUs with only 1\% of their maximum particle load.

The ability to handle particle numbers on the order of 10\textsuperscript{9} on a local workstation enables new possibilities for EL simulations. In addition to excellent scaling, time-to-solution and energy-to-solution have also been shown to decrease significantly.

SCALE-TRACK's capabilities can be enhanced in the future, e.g.~by adding collision models, further boundary interactions, more sophisticated injection models, and an extension to unstructured grids. As it is in principle agnostic to the underlying Euler-solver, coupling to other CFD software than OpenFOAM is also possible. Dynamic load balancing was only implemented as a proof of concept and should also be included. The partitioning would be regularly checked and if the overlap between bounding boxes becomes too large, particles can be exchanged between chunks, or the partitioning can be rearranged.

Regarding verification and validation, a comparison against more challenging test cases would be beneficial to get a better understanding of the influence of time shifted source incorporation. Specifically, steeper temporal or spatial gradients in the fields influencing the particles and, vice versa, in the sources delivered by the particles.

SCALE-TRACK's excellent scaling opens up possibilities that can be useful for a wide range of scientific fields. An obvious example would be high-fidelity simulations of clouds in chambers and in the atmosphere to tackle open questions regarding cloud formation and evolution.

\section*{Acknowledgements}
This project has received funding from the European High-Performance Computing Joint Undertaking (JU) under grant agreement No 101118139. The JU receives support from the European Union’s Horizon Europe Program.

During the preparation of this work, the authors used AI tools of GWDG mbH Göttingen for checking spelling and grammar. The authors reviewed and edited the output as needed and take full responsibility for the content of the published article.

We thank Mira Pöhlker for her valuable feedback on this work.

\section*{Conflict of interest}
The authors have no conflicts to disclose.

\section*{Data Availability Statement}
The SCALE-TRACK source code and a collection of test cases are available in a github repository at \nolinkurl{https://github.com/Wikki-GmbH/SCALE-TRACK}. The underlying OpenFOAM code is available at \nolinkurl{https://gitlab.com/openfoam/core/openfoam/-/tree/OpenFOAM-v2406}. Raw data were generated at MareNostrum5 and LUMI. Derived data supporting the findings of this study are available from the corresponding author upon reasonable request.

\printcredits

\bibliographystyle{cas-model2-names}

\bibliography{SCALE-TRACK_Rev1}

@article{banchelli2025,
title = {{Introducing MareNostrum5: A European pre-exascale energy-efficient system designed to serve a broad spectrum of scientific workloads}},
journal = {Future Generation Computer Systems},
volume = {176},
pages = {108125},
year = {2026},
issn = {0167-739X},
doi = {https://doi.org/10.1016/j.future.2025.108125},
author = {Fabio Banchelli and Marta Garcia-Gasulla and Filippo Mantovani and Joan Vinyals and Josep Pocurull and David Vicente and Beatriz Eguzkitza and Flavio C.C. Galeazzo and Mario C. Acosta and Sergi Girona},
}

@article{bezanson2017,
    title={{Julia: A fresh approach to numerical computing}},
    author={Bezanson, Jeff and Edelman, Alan and Karpinski, Stefan and Shah, Viral B},
    journal={SIAM {R}eview},
    volume={59},
    number={1},
    pages={65--98},
    year={2017},
    publisher={SIAM},
    doi={10.1137/141000671}
}

@article{blume2019,
  title={{3D} simulation of turbulent and cavitating flow for the analysis of primary breakup mechanisms in realistic diesel injection processes},
  author={Blume, Martin and Schwarz, Philip and Rusche, Henrik and Wei{\ss}, Lukas and Wensing, Michael and Skoda, Romuald},
  journal={Atomization and Sprays},
  volume={29},
  number={10},
  year={2019},
  publisher={Begel House Inc.},
  doi={10.1615/AtomizSpr.2020032492}
}

@article{bodenschatz2010,
  title={Can we understand clouds without turbulence?},
  author={Bodenschatz, Eberhard and Malinowski, Szymon P and Shaw, Raymond A and Stratmann, Frank},
  journal={Science},
  volume={327},
  number={5968},
  pages={970--971},
  year={2010},
  publisher={American Association for the Advancement of Science},
  doi={10.1126/science.1185138}
}

@article{chen2025,
  title={{A model intercomparison study of aerosol-cloud-turbulence interactions in a cloud chamber: 1. Model results}},
  author={Chen, Sisi and Krueger, Steven K and Dziekan, Piotr and Enokido, Kotaro and MacMillan, Theodore and Richter, David and Schmalfu{\ss}, Silvio and Shima, Shin-ichiro and Yang, Fan and Anderson, Jesse C and others},
  journal={Journal of Advances in Modeling Earth Systems},
  volume={17},
  number={7},
  pages={e2024MS004562},
  year={2025},
  publisher={Wiley Online Library},
  doi={10.1029/2024MS004562}
}

@article{crowe1977,
  title={The {Particle-Source-In Cell} {(PSI-CELL)} {Model for Gas-Droplet Flows}},
  author={Crowe, C. T. and Sharma, M. P. and Stock, D. E.},
  journal={Journal of Fluids Engineering},
  volume={99},
  number={2},
  pages={325--332},
  year={1977},
  doi={10.1115/1.3448756}
}

@article{dziekan2022,
  title={{University of Warsaw Lagrangian Cloud Model (UWLCM) 2.0: adaptation of a mixed Eulerian--Lagrangian numerical model for heterogeneous computing clusters}},
  author={Dziekan, Piotr and Zmijewski, Piotr},
  journal={Geoscientific Model Development},
  volume={15},
  number={11},
  pages={4489--4501},
  year={2022},
  publisher={Copernicus Publications G{\"o}ttingen, Germany},
  doi={10.5194/gmd-15-4489-2022}
}

@article{ge2020,
  title={{Development of a CPU/GPU portable software library for Lagrangian--Eulerian simulations of liquid sprays}},
  author={Ge, Wenjun and Sankaran, Ramanan and Chen, Jacqueline H},
  journal={International Journal of Multiphase Flow},
  volume={128},
  pages={103293},
  year={2020},
  publisher={Elsevier},
  doi={10.1016/j.ijmultiphaseflow.2020.103293}
}

@article{golshan2023,
  author = {Golshan, S. and Peter Munch and Rene Gassmöller and Martin Kronbichler and Bruno Blais},
  title = {Lethe-DEM: an open-source parallel discrete element solver with load balancing},
  journal = {Computational Particle Mechanics},
  volume = {10},
  number = {1},
  pages = {77-96},
  year = {2023},
  doi = {https://doi.org/10.1007/s40571-022-00478-6}
}

@article{hoefler2010,
  title={Scalable communication protocols for dynamic sparse data exchange},
  author={Hoefler, Torsten and Siebert, Christian and Lumsdaine, Andrew},
  journal={ACM Sigplan Notices},
  volume={45},
  number={5},
  pages={159--168},
  year={2010},
  publisher={ACM New York, NY, USA},
  doi={10.1145/1837853.16934}
}

@article{kemmler2025,
  title={{Efficiency and scalability of fully-resolved fluid-particle simulations on heterogeneous CPU-GPU architectures}},
  author={Kemmler, Samuel and Rettinger, Christoph and R{\"u}de, Ulrich and Cu{\'e}llar, Pablo and K{\"o}stler, Harald},
  journal={The International Journal of High Performance Computing Applications},
  volume={39},
  number={3},
  pages={345--363},
  year={2025},
  publisher={SAGE Publications Sage UK: London, England},
  doi={10.1177/10943420241313}
}

@article{lain2013,
  title={{Characterisation of pneumatic conveying systems using the Euler/Lagrange approach}},
  author={La{\'\i}n, Santiago and Sommerfeld, Martin},
  journal={Powder Technology},
  volume={235},
  pages={764--782},
  year={2013},
  publisher={Elsevier},
  doi={10.1016/j.powtec.2012.11.029}
}

@article{lewy1928,
  title={{{\"U}ber die partiellen Differenzengleichungen der mathematischen Physik}},
  author={Courant, Richard and Friedrichs, K and Lewy, Hans},
  journal={Mathematische Annalen},
  volume={100},
  pages={32--74},
  year={1928},
  doi={10.1007/BF01448839}
}

@article{liu2023,
  title={{A heterogeneous parallel algorithm for Euler-Lagrange simulations of liquid in supersonic flow}},
  author={Liu, Xu and Sun, Mingbo and Wang, Hongbo and Li, Peibo and Wang, Chao and Zhao, Guoyan and Yang, Yixin and Xiong, Dapeng},
  journal={Applied Sciences},
  volume={13},
  number={20},
  pages={11202},
  year={2023},
  publisher={MDPI},
  doi={10.3390/app132011202}
}

@article{marchisio2005,
  title={Solution of population balance equations using the direct quadrature method of moments},
  author={Marchisio, Daniele L and Fox, Rodney O},
  journal={Journal of Aerosol Science},
  volume={36},
  number={1},
  pages={43--73},
  year={2005},
  publisher={Elsevier},
  doi={10.1016/j.jaerosci.2004.07.009}
}

@article{niedermeier2020,
  title={{Characterization and first results from LACIS-T: A moist-air wind tunnel to study aerosol--cloud--turbulence interactions}},
  author={Niedermeier, Dennis and Voigtl{\"a}nder, Jens and Schmalfu{\ss}, Silvio and Busch, Daniel and Schumacher, J{\"o}rg and Shaw, Raymond A and Stratmann, Frank},
  journal={Atmospheric Measurement Techniques},
  volume={13},
  number={4},
  pages={2015--2033},
  year={2020},
  publisher={Copernicus Publications G{\"o}ttingen, Germany},
  doi={10.5194/amt-13-2015-2020}
}

@article{niedermeier2025,
  title={Particle deliquescence in a turbulent humidity field},
  author={Niedermeier, Dennis and Hoffmann, Rasmus and Schmalfuss, Silvio and Frey, Wiebke and Senf, Fabian and Hellmuth, Olaf and P{\"o}hlker, Mira and Stratmann, Frank},
  journal={Aerosol Research Discussions},
  volume={2025},
  pages={1--19},
  year={2025},
  publisher={G{\"o}ttingen, Germany},
  doi={10.5194/ar-3-219-2025}
}

@article{pohlker2023,
  title={Respiratory aerosols and droplets in the transmission of infectious diseases},
  author={P{\"o}hlker, Mira L and P{\"o}hlker, Christopher and Kr{\"u}ger, Ovid O and F{\"o}rster, Jan-David and Berkemeier, Thomas and Elbert, Wolfgang and Fr{\"o}hlich-Nowoisky, Janine and P{\"o}schl, Ulrich and Bagheri, Gholamhossein and Bodenschatz, Eberhard and others},
  journal={Reviews of Modern Physics},
  volume={95},
  number={4},
  pages={045001},
  year={2023},
  publisher={APS},
  doi={10.1103/RevModPhys.95.045001}
}

@article{ruger2000,
  title={{Euler/Lagrange calculations of turbulent sprays: the effect of droplet collisions and coalescence}},
  author={Ruger, M and Hohmann, S and Sommerfeld, Martin and Kohnen, Gangolf},
  journal={Atomization and sprays},
  volume={10},
  number={1},
  year={2000},
  publisher={Begel House Inc.},
  doi={10.1615/AtomizSpr.v10.i1.30}
}

@article{schiller1933,
  title={{Über die grundlegenden Berechnungen bei der Schwerkraftaufbereitung}},
  author={Schiller, L and Naumann, A},
  journal={Zeitschrift des Vereines Deutscher Ingenieure},
  volume={77},
  pages={318--321},
  year={1933}
}

@article{schmalfuss2017,
  title={{Numerical and experimental analysis of Fluid Phase Resonance mixers}},
  author={Schmalfu{\ss}, Silvio and Sommerfeld, Martin},
  journal={Chemical Engineering Science},
  volume={173},
  pages={570--577},
  year={2017},
  publisher={Elsevier},
  doi={10.1016/j.ces.2017.08.012}
}

@article{shaw2025,
  title={Scientific directions for cloud chamber research: instrumentation, modeling, new chambers, and emerging chamber concepts},
  author={Shaw, Raymond A and Chen, Sisi and Freer, Matt and Korolev, Alexei and Krueger, Steve and Murakami, Masataka and Niedermeier, Dennis and Ovchinnikov, Mikhail and Schmalfu{\ss}, Silvio and Tian, Ping and Yang, Fan and Yum, Seong Soo and Zhu, Zeen and Cha, Joo Wan},
  journal={Bulletin of the American Meteorological Society},
  volume={106},
  number={4},
  pages={E770--E781},
  year={2025},
  publisher={American Meteorological Society},
  doi={10.1175/BAMS-D-25-0027.1}
}

@article{sokolichin1997,
  title={{Dynamic numerical simulation of gas-liquid two-phase flows Euler/Euler versus Euler/Lagrange}},
  author={Sokolichin, A and Eigenberger, G and Lapin, A and L{\"u}bert, A},
  journal={Chemical Engineering Science},
  volume={52},
  number={4},
  pages={611--626},
  year={1997},
  publisher={Elsevier},
  doi={10.1016/S0009-2509(96)00425-3}
}

@article{sommerfeld2016,
  title={{Numerical analysis of carrier particle motion in a dry powder inhaler}},
  author={Sommerfeld, Martin and Schmalfu{\ss}, Silvio},
  journal={Journal of Fluids Engineering},
  volume={138},
  number={4},
  pages={041308},
  year={2016},
  publisher={American Society of Mechanical Engineers},
  doi={10.1115/1.4031693}
}

@article{sommerfeld2020,
  title={{Analysis and optimisation of particle mixing performance in fluid phase resonance mixers based on Euler/Lagrange calculations}},
  author={Sommerfeld, Martin and Schmalfu{\ss}, Silvio},
  journal={Advanced Powder Technology},
  volume={31},
  number={1},
  pages={139--157},
  year={2020},
  publisher={Elsevier},
  doi={10.1016/j.apt.2019.10.006}
}

@article{stock2023,
  author = {A. Stock and G. Lartigue and V. Moureau},
  title = {Diffusive orthogonal load balancing for Euler–Lagrange simulations},
  journal = {International Journal for Numerical Methods in Fluids},
  volume = {95},
  number = {8},
  pages = {1220-1239},
  doi = {https://doi.org/10.1002/fld.5191},
  year = {2023}
}

@article{sweet2018,
  title={{GPU acceleration of Eulerian--Lagrangian particle-laden turbulent flow simulations}},
  author={Sweet, James and Richter, David H and Thain, Douglas},
  journal={International Journal of Multiphase Flow},
  volume={99},
  pages={437--445},
  year={2018},
  publisher={Elsevier},
  doi={10.1016/j.ijmultiphaseflow.2017.11.010}
}

@article{thari2022,
author = {A. Thari and M. Staufer and G.J. Page},
title = {Asynchronous task based Eulerian-Lagrangian parallel solver for combustion applications},
journal = {Journal of Computational Physics},
volume = {458},
pages = {111103},
year = {2022},
doi = {https://doi.org/10.1016/j.jcp.2022.111103}
}

@article{wang2024,
  author={Wang, Aaron and Ovchinnikov, Mikhail and Yang, Fan and Schmalfu{\ss}, Silvio and Shaw, Raymond A},
  title={Designing a convection-cloud chamber for collision-coalescence using large-eddy simulation with bin microphysics},
  journal={Journal of Advances in Modeling Earth Systems},
  volume={16},
  number={1},
  pages={e2023MS003734},
  year={2024},
  publisher={Wiley Online Library},
  doi={10.1029/2023MS003734}
}

@article{wang2022,
  title={{An GPU-accelerated particle tracking method for Eulerian--Lagrangian simulations using hardware ray tracing cores}},
  author={Wang, Bin and Wald, Ingo and Morrical, Nate and Usher, Will and Mu, Lin and Thompson, Karsten and Hughes, Richard},
  journal={Computer Physics Communications},
  volume={271},
  pages={108221},
  year={2022},
  publisher={Elsevier},
  doi={10.1016/j.cpc.2021.108221}
}

@article{xu2012,
  title={{Discrete particle simulation of gas--solid two-phase flows with multi-scale CPU--GPU hybrid computation}},
  author={Xu, Ming and Chen, Feiguo and Liu, Xinhua and Ge, Wei and Li, Jinghai},
  journal={Chemical engineering journal},
  volume={207},
  pages={746--757},
  year={2012},
  publisher={Elsevier},
  doi={10.1016/j.cej.2012.07.049}
}

@book{michaelides2017,
  title={{Multiphase Flow Handbook}},
  author={Michaelides, Efstathios E. and Crowe, Clayton T. and Schwarzkopf, John D.},
  year={2017},
  publisher={CRC {Press}}
}

@inproceedings{dennis2024,
  title={{A portable and efficient Lagrangian particle capability for idealized atmospheric phenomena}},
  author={Dennis, John and Sun, Jian and Voelz, Sheri and Bryan, George and Richter, David H},
  booktitle={Proceedings of the Platform for Advanced Scientific Computing Conference},
  pages={1--11},
  year={2024},
  doi={10.1145/3659914.3659940}
}

@inproceedings{galeazzo2024,
  title={{Understanding superlinear speedup in current HPC architectures}},
  author={Galeazzo, Flavio Cesar Cunha and Wei{\ss}, R Gregor and Lesnik, Sergey and Rusche, Henrik and Ruopp, Andreas},
  booktitle={IOP Conference Series: Materials Science and Engineering},
  volume={1312},
  pages={012009},
  year={2024},
  organization={IOP Publishing},
  doi={10.1088/1757-899X/1312/1/012009}
}

@inproceedings{muralikrishnan2024,
  title={Scaling and performance portability of the particle-in-cell scheme for plasma physics applications through mini-apps targeting exascale architectures},
  author={Muralikrishnan, Sriramkrishnan and Frey, Matthias and Vinciguerra, Alessandro and Ligotino, Michael and Cerfon, Antoine J and Stoyanov, Miroslav and Gayatri, Rahulkumar and Adelmann, Andreas},
  booktitle={Proceedings of the 2024 SIAM Conference on Parallel Processing for Scientific Computing (PP)},
  pages={26--38},
  year={2024},
  organization={SIAM},
  doi={10.1137/1.9781611977967.3}
}

@inproceedings{shaw2025design,
    author = {{Shaw}, Raymond A. and {Anderson}, Jesse C. and {Bakri}, Zaid and {Bench}, Jim and {Bewley}, Gregory and {Blankenship}, Vine and {Cantrell}, Will and {Chandrakar}, Kamal Kant and {Fahandezh sadi}, Hamed and {Flagan}, Richard and {Fuentes}, Jose D. and {Gogos}, George and {Healey}, Ryan and {Iyer}, Kartik and {Kaufman}, Graham and {Kim}, Kwonil and {Kollias}, Pavlos and {Krueger}, Steven K. and {Lamer}, Katia and {Luke}, Edward and {Mazzoleni}, Claudio and {McComiskey}, Allison and {Megaridis}, Constantine and {Mukhopadhyay}, Arani and {Niedermeier}, Dennis and {Oue}, Mariko and {Ovchinnikov}, Mikhail and {Pal}, Anish and {Papailias}, Ilias and {Rajagopal}, Manikandan and {Ren}, Yangze and {Salari}, Ali and {Schmalfuss}, Silvio and {Sedlacek}, Arthur and {Shilling}, John E. and {Shrivastava}, ManishKumar and {Singh}, Suryadev Pratap and {Stratmann}, Frank and {Sua}, Yong Meng and {Thomas}, Lois and {Sulaiman}, Swafuva Varappillikudy and {Wang}, Aaron and {Yang}, Fan and {Yeom}, Jaemin and {Zawadowicz}, Maria Anna and {Zhang}, Jie and {Zhu}, Zeen and {Zuhlke}, Craig},
    title = "{Design of a Convection-Cloud Chamber for Exploring Condensational and Collisional Growth Processes Relevant to Precipitation Formation}",
    booktitle = {105th Annual AMS Meeting},
    year = 2025,
    series = {American Meteorological Society Meeting Abstracts},
    volume = {105},
    month = jan,
    eid = {457167},
    pages = {457167},
    adsurl = {https://ui.adsabs.harvard.edu/abs/2025AMS...10557167S}
}

@misc{of2406,
  author = {{OpenCFD Ltd}},
  title = {{OpenFOAM-v2406}},
  howpublished = "\url{https://www.openfoam.com/news/main-news/openfoam-v2406}",
  year = {2024}, 
  note = "[Online; accessed 13-Nov-2025]"
}

@phdthesis{jaiswal2025,
  title={{Advanced Numerical Modelling of Particle-laden Turbulent Flow with Emphasis on Fine Sediment Infiltration into Static Gravel Bed}},
  author={Jaiswal, Atul},
  year={2025},
  school={Technical University Munich},
  url={https://mediatum.ub.tum.de/1755170}
}

@phdthesis{martinez2009,
  title={{Development and validation of the Euler-Lagrange formulation on a parallel and unstructured solver for large-eddy simulation}},
  author={Martinez, Marta Garc{\'\i}a},
  year={2009},
  school={Institut National Polytechnique de Toulouse-INPT},
  url={https://theses.hal.science/tel-00414067/}
}

@phdthesis{schmalfuss2023,
  title={{Transportvorg{\"a}nge in Fluidphasenresonanzmischern}},
  author={Schmalfu{\ss}, Silvio},
  year={2023},
  school={Otto-von-Guericke-University Magdeburg},
  doi={10.25673/108451}
}

\end{document}